\documentclass[a4paper,12pt,english]{article}
\usepackage{amsfonts}
\usepackage{graphicx}
\usepackage{amsmath}
\usepackage{color}

\newcommand{\be}{\begin{equation}}
\newcommand{\ee}{\end{equation}}
\begin{document}


\begin{titlepage}
\begin{center}

\noindent{{\LARGE{The mass of a Lifshitz black hole}}}

\smallskip
\smallskip

\smallskip
\smallskip

\smallskip
\smallskip
\smallskip
\smallskip
\noindent{\large{Gaston Giribet$^1$, Edmundo Lavia$^{1,2}$}}

\smallskip
\smallskip

\smallskip
\smallskip

\smallskip
\smallskip

\centerline{$^1$ Physics Department, University of Buenos Aires FCEyN-UBA and IFIBA-CONICET}
\centerline{{\it Ciudad Universitaria, pabell\'on 1, 1428, Buenos Aires, Argentina.}}

\smallskip
\smallskip

\centerline{$^2$ Acoustic Propagation Department. Argentinian Navy Research Office (DIIV)}
\centerline{{\it Laprida 555, 1638 Vicente L\'opez, Buenos Aires, Argentina.}}

\end{center}

\bigskip

\bigskip

\bigskip

\bigskip

\begin{abstract}
It is well known that massive 3D gravity admits solutions that describe Lifshitz black holes as those considered in non-relativistic holography. However, the determination of the mass of such black holes remained unclear as many different results were reported in the literature presenting discrepancies. Here, by using a robust method that permits to tackle the problem in the strong field regime, we determine the correct mass of the Lifshitz black hole of the higher-derivative massive gravity and compare it with other results obtained by different methods. Positivity of the mass spectrum demands an odd normalization of the gravity action. In spite of this fact, the result turns out to be consistent with computations inspired in holography. 
\end{abstract}
\end{titlepage}




\section{Introduction}

The holographic description of $d$-dimensional strongly correlated, non-relativistic systems with anisotropic scale invariance and no Galilean symmetry has been studied long time ago \cite{Kachru}. This consists of a geometrical realization that involves a especial type of static $d+1$-dimensional spacetimes, known as Lifshitz metrics. These read
\begin{equation}
ds^2= -\frac{r^{2z}}{\ell^{2z}}\, dt^2 + \frac{\ell^2}{r^2}\, dr^2 + \frac{r^2}{\ell^2}\, dx^2\label{Lifa}
\end{equation}
with $t\in \mathbb{R}$, $r\in \mathbb{R}_{>0}$, and $dx^2$ being the flat metric on $\mathbb{R}^{d-1}$; here, we will consider $d=2$, so $x\in \mathbb{R}$. The parameter $z\in \mathbb{R}$ is the so-called dynamical exponent, and $\ell$ is a length scale associated to the spacetime curvature. Despite having finite scalar curvature invariants, the spacetimes (\ref{Lifa}) with $0\neq z \neq 1$ are singular; they are geodesically incomplete for timelike geodesics ending at $r=0$. For $z=1$, in contrast, the metric (\ref{Lifa}) is locally equivalent to AdS$_3$ spacetime, and the case $z=0$ corresponds to the space product $\mathbb{R}\, \times $ AdS$_2$. 
For $z$ generic, spacetimes (\ref{Lifa}) enjoy scale invariance
\begin{equation}
t\to e^{z\sigma}t \ , \ \ \ \ r\to e^{-\sigma}r \ , \ \ \ \ x \to e^{\sigma}x \, ,\label{Uscaling}
\end{equation}
with $\sigma$ being an arbitrary constant. This scaling symmetry, together with the translations in $t$ and $x$, generate the full isometry group. The cases $z=0$ and $z=1$ are of course especial, having 4 and 6 Killing vectors and generating the groups $\mathbb{R} \times SL(2,\mathbb{R})$ and $SL(2,\mathbb{R})\times SL(2,\mathbb{R})$, respectively. For $z$ arbitrary, the Killing vectors are
\begin{equation}
H=\partial_t \ , \ \ \ \ P=\partial_{x} \ , \ \ \ \ D=zt\partial_t - r\partial_r + x \partial_{x }\, ,
\end{equation}
and generate the nilpotent isometry algebra
\begin{equation}
[P,H]=0 \ , \ \ \ \ [D,P]=P \ , \ \ \ \ [D,H]=z H .
\end{equation}

The geometric configuration that would holographically describe $2$-dimensional Lifshitz-type systems with dynamical exponent $z$ at finite temperature are $3$-dimensional black holes that asymptote (\ref{Lifa}) at large $r$. This motivates the search for sensible models that admit such black holes as exact solutions. This is actually a hard problem due to the validity of Birkhoff-type theorems in a large variety of systems, precluding the existence of static black hole configurations of the type required. This is the reason why the construction of asymptotically Lifshitz black holes typically involves the introduction of exotic matter content or non-minimal couplings to the gravity sector. However, it turns out that, in 3 dimensions, there exists a remarkably simple model admitting Lifshitz black holes. This is given by the massive deformation of 3-dimensional Einstein theory with no additional fields. It was shown in \cite{Lifshitz} that, if one considers the parity-even massive 3D gravity proposed in \cite{NMG}, a static Lifshitz black hole solution with dynamical exponent $z=3$ can be analytically constructed. While other models admitting Lifshitz black holes are known in 3 dimensions, these either include additional fields \cite{otroGoya, otroEloy} or exotic gravity field equations \cite{otroYo}. This makes the simple instances of Lifshitz black hole scarce. An example of this is massive gravity itself, where it has been proven \cite{Bircof} that such static black holes only exist for $z=1$ and $z=3$. This is why the solution of \cite{Lifshitz} is particularly interesting.

Simple scaling arguments show that the mass of the $z=3$ Lifshitz black hole of 3D massive gravity -- see (\ref{Lifshitz}) below-- takes the form
\begin{equation}
M = \eta \,  \frac{ L \, r_+^4}{2\pi G\ell^5}\label{LaMasa}
\end{equation}
where $\eta $ is a dimensionless coefficient, $G$ is the Newton constant, $r_+$ is the horizon radius, $L$ is the length of the segment in which $x$ takes values, and $\ell$ is the length scale that appears in (\ref{Lifa}) and which relates to the scalar curvature of the black hole as follows
\begin{equation}
R=\frac{2}{\ell^2}\Big( -13 + 4\, \frac{r_+^2}{r^2}\Big)\, .\label{Richi}
\end{equation} 
It is usual to consider the black hole solution with the coordinate $x\equiv \varphi \ell$ being periodic with period $2\pi \ell$. This of course breaks the scaling symmetry, making the isometry group to be $\mathbb{R} \times SO(2)$ even asymptotically. Here, we will consider $\varphi \in [0,2\pi]$, namely $L=2\pi \ell $.

In the literature, different authors, using different methods to compute the conserved charges in higher-curvature theories, have arrived to different results for the value of $\eta $ in (\ref{LaMasa}). This raises the question about the correct value of the gravitational energy in the Lifshitz spacetime. The discrepancy among different authors can be explained by the difficulty of computing conserved charges for solutions of higher-derivative theories that exhibit non-standard asymptotics. The problem with this is twofold: firstly, there exists an ambiguity in the choice of counterterms when higher-derivatives terms are present; secondly, the {\it empty} Lifshitz spacetime is actually singular, what makes the problem of identifying the correct reference background less clear. This results in that not all the machinery that we at hand when dealing with asymptotically maximally symmetric spacetimes can actually be successfully applied to the case of Lifshitz spacetimes. This led the people to consider many different methods, with different degree of success. In \cite{Cai:2009ac}, for example, the author considered the Wald formula to compute the entropy and inferred the mass from the first law of black hole mechanics, having found (\ref{LaMasa}) with $\eta = -1/4$. In \cite{Myung:2009up}, in contrast, the authors considered a method involving dimensional reduction and found $\eta = 1/16$. In \cite{Hohm:2010jc}, the value $\eta = -1/4$ was found by defining a holographic stress-tensor and computing the quasi-local energy. In \cite{Devecioglu:2010sf}, the authors adapted the Abbott-Deser-Tekin (ADT) approach \cite{ADT} to spaces with non-constant curvature and found $\eta = 7/8$. In \cite{Gonzalez:2011nz}, the authors made a very interesting analysis of the Lifshitz black hole thermodynamics and showed that this was consistent with $|\eta |= 1/4$. The value $\eta = +1/4$ was found in \cite{Gim:2014nba} considering another adaptation of ADT. Here, by considering a robust method that dispenses with the analysis of the large-radius asymptotia and permits to deal with the problem in the strong field regime, we will show that the correct value for the mass of the $z=3$ Lifshitz black hole of the massive 3-dimensional gravity is (\ref{LaMasa}) with $\eta=-1/4$. In particular, this implies that the mass of the black hole is negative for positive $G$ and, therefore, as usual in massive 3D gravity, one needs to consider the {\it wrong} sign of the Newton constant in order to make sense out of the Lifshitz background.

\section{Massive 3D gravity}

Let us begin by reviewing the 3-dimensional massive gravity theory and its solutions. The action of the theory is
\begin{equation}
I=\frac{1}{16\pi G}\int d^3x \, \sqrt{-g}\, \Big[ R-2 \lambda-\frac{1}{m^2} \Big( R_{\mu \nu}R^{\mu \nu}-\frac{3}{8}R^2\Big) \Big] \, . \label{ActionNMG}
\end{equation}
This theory exhibits two local degrees of freedom organized in a way that there is a massive spin-2 mode of mass $m$. At linearized level, and around maximally symmetric spaces, the theory coincides with the spin-2 Fierz-Pauli theory \cite{NMG}. This implies that action (\ref{ActionNMG}) describes a ghost-free theory. At full non-linear level, the field equations take the form 
\begin{equation}\label{eom}
R_{\mu \nu}-\frac 12 Rg_{\mu \nu }+\lambda g_{\mu \nu}-\frac{1}{2m^2}K_{\mu \nu}=0,
\end{equation}
with
\begin{equation}
K_{\mu \nu}=2\nabla^2R_{\mu\nu}-\frac{1}{2}(\nabla_\mu \nabla_\nu R+g_{\mu \nu}\nabla^2 R)-8 R_{\mu \rho}R^{\rho}_{\,\,\nu}+\frac{9}{2}RR_{\mu\nu}+\frac{1}{8}g_{\mu\nu}\left(24R^{\alpha \beta}R_{\alpha \beta}-13R^2\right).\label{JKL}
\end{equation} 
In the infinite mass limit, $m^2\to \infty$, where the local degrees of freedom decouple, the theory reduces to 3-dimensional Einstein gravity.

Being a quadratic-curvature theory, for generic values of $\lambda$ and $m$ the field equations (\ref{eom})-(\ref{JKL}) may admit two maximally symmetric solutions. That is to say, generically there exist two values of the effective cosmological constant; these are
\begin{equation}
\Lambda_{\pm } = 2m^2 \pm 2m^2\sqrt{1-\frac{\lambda }{m^2} }, \label{Opa}
\end{equation}
assuming $m^2\geq \lambda$. This means that the theory has two natural vacua, which can be either Minkowski or (A)dS spaces, depending on the range of parameters. The effective cosmological constants (\ref{Opa}) give the curvature radius of the solution $\ell=1/\sqrt{-\Lambda_{\pm}}$; $\ell^2>0$ for AdS$_3$. This is equivalent to say
\begin{equation}
\lambda = -\frac{1}{\ell^2}\Big( 1+ \frac{1}{4m^2\ell^2}\Big)\, . \label{Opaz}
\end{equation}
For $\Lambda _{\pm }<0$, the theory admits asymptotically AdS$_3$ solutions, including Ba\~nados-Teitelboim-Zanelli (BTZ) black holes \cite{BTZ} and other interesting solutions \cite{Bergshoeff:2009aq, Oliva:2009ip}. The theory also admits solutions (\ref{Lifa}) for arbitrary $z$ provided the coupling constants take the values
\begin{equation}
m^2\ell^2=-\frac 12 (z^2-3z+1) \ , \ \ \ \ \lambda \ell^2=-\frac 12 (z^2+z+1)  \, ; 
\end{equation}
which in particular demands $\lambda \ell^2<0$. 

\section{Lifshitz black hole}

A remarkable surprise occurs at $z=3$, where the theory admits an extra static black hole solution \cite{Lifshitz}. This happens on a curve in the parameter space where
\begin{equation}
\lambda  = 13m^2 \, . \label{otropunto}
\end{equation}
On this curve, the following black hole solution exists
\begin{equation}
ds^2 = - \frac{r^4}{{{\ell}}^4} \Big(\frac{r^2-r^2_+}{{{\ell}}^2} \Big) \, dt^2 + \Big(\frac{{{\ell}}^2}{r^2-r^2_+} \Big)\, dr^2 + {r^2} \, d{{\varphi }}^2, \label{Lifshitz}
\end{equation}
where $t\in \mathbb{R}$ and $r\in \mathbb{R}_{>0}$. We consider $\varphi $ periodic with period $2\pi $. $r_+$ is an integration constant that represents the horizon location, and ${{\ell}}$ is given by ${{\ell}}^2 = -{1}/({2m^2}) = -{13}/({2\lambda })$.

Metric (\ref{Lifshitz}) is not locally conformally flat, so it is neither a solution of Einstein theory nor of conformal gravity. Furthermore, it is not a solution of the parity-odd Topologically Massive Gravity model. It has isometry group $\mathbb{R}\times SO(2)$, generated by the Killing vectors $\partial_t$, $\partial_{{{\varphi }}}$. The spacetime described by (\ref{Lifshitz}) exhibits a regular event horizon at $r=r_+$, provided $r_+>0$. This horizon shields a curvature singularity that exists at $r=0$; there, the Ricci scalar invariant (\ref{Richi}) together with other invariants like $R_{\mu\nu}R^{\mu\nu}$ diverge. When $r_+=0$, metric (\ref{Lifshitz}) reduces to the Lifshitz space (\ref{Lifa}) with $z=3$. For generic values of $r_+$, the metric still asymptotes Lifshitz space (\ref{Lifa}) with $z=3$ at large $r$, meaning that it is asymptotically, locally invariant under the rescaling $t\to e^{3\sigma } t$, $r\to e^{-\sigma } r$, $\varphi\to e^{\sigma } \varphi $. Actually, the solution also exhibits such a scaling symmetry at finite $r$ provided, in addition to rescaling the coordinates, one also rescales the parameter as $r_+\to e^{-\sigma}r_+$. This leaves the black hole metric invariant. On the one hand, this is consistent with the fact that all the curvature invariants of the Lifshitz black hole depend only on the ratio $r_+^2/r^2$. On the other hand, this provides us with an argument to anticipate the functional dependence of the mass, this being given in (\ref{LaMasa}). We will compute the mass explicitly below.

\section{Conserved charges}


Boundary charges in $d$-dimensional theory of gravity, as well as in a $d$-dimensional gauge theory, are usually understood as integrals of $(d-2)$-form potentials of the free theory, obtained this by linearizing the solution around an appropriate background configuration. These conserved $(d-2)$-forms are in correspondence with the so-called reducibility parameters of the background geometry. In \cite{Glenn2}, a closed $(d-2)$-form for the fully interacting theory has been constructed. It admits a closed form in terms of a one parameter family of solutions to the fully interacting theory admitting one such reducibility parameter. Here, we will consider the method of \cite{Glenn, Glenn2} to compute the charges. This method is fully constructive and robust, and it can be easily adapted to the massive deformation of gravity theory in 3 dimensions. Applying it in the deep bulk region, we will compute the mass of the Lifshitz black hole for the fully interacting theory. 

The expression of the functional variation of the conserved charge associated to the Killing vector $\xi $ is
\begin{equation}\label{lacharga}
\delta Q[\xi ; g,\delta g]=\frac{1}{16\pi G}\int_0^{2\pi} d\varphi\, \sqrt{-g} \,\epsilon_{\mu \nu \varphi } \, k^{\mu \nu}_\xi[g,\delta g] ,  
\end{equation}
where $\delta g_{\mu \nu } = h_{\mu \nu }$ is a perturbation around a solution $g_{\mu \nu }$, and where $k^{\mu \nu}$ is the surface 1-form potential. In the case of the massive 3D gravity, this form is given by three different contributions, namely 
\begin{equation}
k^{\mu \nu} = k^{\mu \nu}_{(0,1)}-\frac{1}{m^2} \, k^{\mu \nu}_{(0,2)}+\frac {3}{8m^2} \, k^{\mu \nu}_{(1,1)}\,  , 
\end{equation}
the first contribution being the one coming from the Einstein-Hilbert term:
\begin{equation}
k^{\mu \nu}_{(0,1)}=\xi_\alpha \nabla^{[\mu}h^{\nu]\alpha}-\xi^{[\mu}\nabla_\alpha h^{\nu]\alpha}-h^{\alpha[\mu}\nabla_\alpha \xi^{\nu]}+\xi^{[\mu}\nabla^{\nu]}h+\frac{1}{2}h\nabla^{[\mu}\xi^{\nu]}.
\end{equation}
The other two contributions come from the higher-derivative terms in the action (\ref{ActionNMG}); they are \cite{Nam}
\begin{eqnarray}
k^{\mu \nu}_{(0,2)}&=&\nabla^2k^{\mu \nu}_{(0,1)}+\frac{1}{2}k^{\mu \nu}_{(1,1)}-2k^{\alpha [\mu }_{(0,1)}R^{\nu]}_\alpha-2\nabla^\alpha\xi^\beta \nabla_\alpha \nabla^{[\mu}h^{\nu]}_\beta-4\xi^\alpha R_{\alpha \beta}\nabla^{[\mu}h^{\nu]\beta}\nonumber \\
&&+2\xi^{[\mu}R^{\nu]}_\alpha \nabla_\beta h^{\alpha \beta} +2\xi_\alpha R^{\alpha[\mu}\nabla_\beta h^{\nu]\beta}+2\xi^\alpha h^{\beta[\mu}\nabla_\beta R^{\nu]}_\alpha+2 h^{\alpha \beta} \xi^{[\mu} \nabla_\alpha R^{\nu]}_\beta \label{Kopp} \\
&&-(\delta R +2 R^{\alpha \beta}h_{\alpha \beta})\nabla^{[\mu}\xi^{\nu]}-3\xi^{\alpha}R_\alpha^{[\mu}\nabla^{\nu]}h-\xi^{[\mu}R^{\nu]\alpha}\nabla_\alpha h -Rh^{[\mu}_\alpha \nabla^{\nu]}\xi^\alpha,\nonumber 
\end{eqnarray}
and
\begin{eqnarray}\label{Kop}
k^{\mu \nu}_{(1,1)}=2R\,k^{\mu \nu}_{(0,1)}+4\xi^{[\mu}\nabla^{\nu]}\delta R+2\delta R \nabla^{[\mu}\xi^{\nu]}-2\xi^{[\mu}h^{\nu]\alpha}\nabla_\alpha R,
\end{eqnarray}
with $h_{\mu\nu }=\delta g_{\mu\nu }$, $\delta R=-R^{\alpha \beta}h_{\alpha \beta}+\nabla^\alpha \nabla^\beta h_{\alpha \beta}-\nabla^2 h$, and $h=h_{\mu}^{\,\mu }$.

As said, we will address the computation of the charges in the region of the space where the theory is fully interacting. To do so, we find convenient to take the phase space of metric in their near-horizon form. We will consider the near horizon boundary conditions studied in \cite{DGGP}; namely, near the horizon consider the metric in the form
\begin{equation}\label{ds2}
ds^2\, =\, g_{\mu\nu }\, dx^{\mu} dx^{\nu}\, =\, f(v,\varphi)\, dv^2 -2 k(v,\varphi)\, dv d\rho + 2h(v,\varphi)\, dv d\varphi+R^2(v,\varphi)\, d\varphi^2,
\end{equation}
where $v\in \mathbb{R}$, $\rho\geq 0$, and $\varphi \in [0,2\pi ]$ with period $2\pi $. The metric functions are of the form
\begin{equation}
\label{boundaryconditions}
\begin{split}
f(v,\varphi)&= -2\kappa \,\rho + g_{vv}^{(2)}(\varphi) \,\rho^2+\, ... \\
k(v,\varphi)&=1 -g_{v\rho }^{(2)} (\varphi )\, \rho ^2+ \, ... \\
h(v,\varphi)&= g_{v\varphi }^{(1)}(\varphi)\,\rho+g_{v \varphi }^{(2)}(\varphi)\, \rho^2+\, ... \\
R^2(v,\varphi)&=g_{\varphi \varphi }^{(0)}(\varphi)+ g_{\varphi \varphi }^{(1)}(\varphi)\, \rho + g_{\varphi \varphi }^{(2)}(\varphi, v)\, \rho^2 +\, ...
\end{split}
\end{equation}
where the ellipsis stand for functions of $v$ and $\varphi $ that vanish at least as fast as ${\mathcal O}(\rho^3)$ near the surface $\rho =0$ where the horizon is located. Notation is such that $g_{\mu \nu }^{(n)}$ are the $\rho $-independent functions that accompany the order ${\mathcal O}(\rho^n)$ in the power expansion. In the expressions above, $g_{\varphi \varphi }^{(0)}$, $g_{v\varphi }^{(1)}$, $g_{\varphi \varphi }^{(1)}$, $g_{vv}^{(2)}$, $g_{v \varphi }^{(2)}$, and $g_{v\rho }^{(2)} $ are arbitrary functions of the coordinate $\varphi $, while $\kappa =-\frac 12 g_{vv}^{(0)}$ corresponds to the surface gravity at the horizon and thus is constant. We have also fixed $g_{v\rho }^{(1)}=0$, and we could have even set the gauge $g_{v\rho }=1$ together with $g_{\rho \rho }=0$.

As a first check that this way of computing the charges actually works, let us illustrate the calculation considering the BTZ black hole. We evaluate (\ref{lacharga}) for the Killing vector $\xi = \partial_v$ and realize the functional variation by varying the parameter $r_+$; that is, we perform $r_+\to r_+ + \delta r_+$. This induces a variation of the near horizon form of the BTZ metric $g_{\mu \nu }\to g_{\mu \nu }+\delta g_{\mu \nu }$, with
\begin{equation}
\delta g_{\mu \nu } = - \frac{2}{\ell^2}\rho \, \delta r_+\, dv^2 + {2}(r_++\rho ) \delta r_+ \, d\varphi^2\, ,
\end{equation}
and after integrating we find
\begin{equation}\label{jjjjj}
Q[\partial_v ; g, \delta g] = \frac{r^2_+}{8G\ell^2} \Big( 1- \frac{1}{2m^2 \ell^2}\Big)\, ,
\end{equation}
which is actually the correct result for the mass of the BTZ black hole in the massive gravity theory. In addition, in order to check this method, we can try to follow the same steps to compute the mass of the generalization of the BTZ black hole that, for massive gravity theory, was found in \cite{Bergshoeff:2009aq, Oliva:2009ip}; see Eqs. (24)-(25) in the latter reference. This black hole, which only exists when ${2m^2 \ell^2}=-1$, has non-constant curvature, is asymptotically AdS$_3$ in a way that is weaker than the standard Brown-Henneaux boundary conditions, and presents two horizons; let us denote $r_{\pm}$ the location of the horizons and $\delta r_{\pm}$ their independent variations. This yields
\begin{equation}
\delta g_{\mu \nu } = - \frac{1}{\ell^2}\rho \, (\delta r_+-\delta r_-)\, dv^2 + {2}(r_++\rho ) \delta r_+ \, d\varphi^2\, ,
\end{equation}
which gives
\begin{equation}\label{jjjjji}
Q[\partial_v ; g, \delta g] = \frac{(r_+-r_-)^2}{16G\ell^2}  .
\end{equation}
This actually coincides with the correct value of the mass; see Eq. (8) in \cite{Giribet:2009qz}; see also (12) in \cite{conJulio}, cf. Eq. (49) therein. In the particular case $r_-=-r_+$ the solution reduces to the static BTZ black hole, and in that case (\ref{jjjjji}) reduces to (\ref{jjjjj}) for ${2m^2 \ell^2}=-1$. This indicates that the method of computing the mass from the near horizon charges is working perfectly, even in the case of black holes with non-constant curvature. At this point, one might wonder why this near horizon computation is giving the correct value of the mass and not, as in \cite{DGGP}, the product between the Hawking temperature and the Bekenstein-Hawking entropy, cf. \cite{DGGP2, conJulio}. The answer is that, while the near horizon boundary conditions considered here are exactly the same as in \cite{DGGP}, the way in which we implement the functional variation here is different: Here, we do not consider variations in the space of metrics that keep the horizon temperature constant, but we consider arbitrary variations in a one- or two-parameters family. In other words, $\delta g$ in (\ref{lacharga}) here generically yields $\delta g^{(0)}_{vv}\neq 0$. As a result, we correctly reproduce the black hole mass from the near horizon computation, with the appropriate numerical factor.

In the case of the BTZ black hole, the same result (\ref{jjjjj}) can be obtained by resorting to the ADT method, which amounts to consider linearized solutions around the AdS$_3$ vacuum in the asymptotic, near boundary region. However, in the case of the $z=3$ Lifshitz black hole, the method that resorts to the linearization of the metric in the large-$r$ region does not lead to the correct result for the mass. The reason why it happens has been explained in \cite{Nam}. In that case, the computation yields
\begin{equation}
\bar{M} = \frac{7}{4}\frac{r^4_+}{G\ell^4}\, .
\end{equation}
We confirm this output, which is not the correct result for the $z=3$ Lifshitz black hole. The correct value for the mass of the latter can be obtained as we did above for the case of the $z=1$ solutions. However, this would first require to put the solution (\ref{Lifshitz}) in the near horizon form (\ref{ds2})-(\ref{boundaryconditions}). To achieve so, we define coordinates
\begin{equation}\label{Oot}
v = t - {{\ell}}^4\, \int \frac{dr}{r^2 (r^2-r^2_+)} \ ,  \ \ \ \ \ \rho = \frac{r^3 - r_+^3}{3{{\ell}}^2}\, .
\end{equation}
We observe that $\rho = 0$ at the horizon $r=r_+$, and it holds that
\begin{equation}
r-r_+\simeq \frac{{{\ell}}^2}{r_+^2}\rho + \mathcal{O}(\rho^2)
\end{equation}
for small $\rho$. The change of variable (\ref{Oot}) suffices to put metric (\ref{Lifshitz}) in the form (\ref{ds2})-(\ref{boundaryconditions}) with 
\begin{equation}\label{quiza}
g_{vv}^{(0)} = -\frac{2r_+^3}{{{\ell}}^4} \ , \ \ \ \ \ g_{vv}^{(2)}= -\frac{7}{\ell^2} \ ,\ \ \ \ \ g_{\varphi \varphi }^{(0)} = {r_+^2} \ ,\ \ \ \ \ g_{\varphi \varphi }^{(1)} = \frac{2\ell^2}{r_+},
\end{equation}
which in particular yields the surface gravity $\kappa = {r_+^3}/{{{\ell}}^4}$.

Now, we are ready to evaluate (\ref{lacharga}) for the Killing vector $\xi = \partial_v$ and realize the functional variation by varying the parameter $\delta r_+$. This yields the metric variation $g_{\mu \nu }\to g_{\mu \nu }+\delta g_{\mu \nu }$ with
\begin{equation}
\delta g_{\mu \nu } = - \frac{6r_+^2}{\ell^4}\rho \, \delta r_+\, dv^2 + \frac{2}{r_+^2}(r_+^3-\ell^2\rho ) \delta r_+ \, d\varphi^2\, .
\end{equation}
And, finally, we obtain
\begin{equation}
Q[\partial_v ; g, \delta g] = -\frac{r^4_+}{4G\ell^4}\frac{L}{2\pi \ell }
\end{equation}
which is the correct result for the mass; that is, $\eta =-1/4$. The factor $L/(2\pi \ell )$ in this expression comes from the integration on the coordinate $x=\varphi \ell $, and it is 1 for the case $\varphi $ has a period $2\pi $.

\section{Conclusions}

In summary, we conclude that the mass, the entropy and the temperature of the $z=3$ black hole solution are given by
\begin{equation}
M = -\frac{r_+^4}{4G\ell^4}, \ \ \ \ \ S= -\frac{2\pi r_+}{\hbar\, G}, \ \ \ \ \ T=\frac{\hbar\, r_+^3}{2\pi \ell^4}\, ,\label{NR}
\end{equation}
respectively. While the entropy can be computed by the Wald formula, the temperature follows from the standard geometrical methods. These quantities satisfy the first principle $dM=TdS$ and a Smarr type formula $M=\frac{1}{4}TS$. Notice that, despite being a solution of a higher-curvature theory, the Lifshitz black hole happens to satisfy the area law $S\propto 2\pi r_+/G$, though with a special factor. Both the mass and the entropy turn out to be negative, so the change $G\to -G$ is needed for making sense out of the theory around this background. 

One may wonder what happens in the case of stationary, non-static black holes. In the case of asymptotically AdS$_3$ rotating black holes, a near-horizon computation in massive 3D gravity was done in \cite{conJulio}. In the case of the rotating version of (\ref{Lifshitz}), such solution actually exists \cite{Stationary} and can be analytically constructed by an improper boost acting on the static metric; however, the resulting spacetime happens not to be asymptotically Lifshitz.

Before concluding, it would be interesting to compare our result with those of the literature and to explain the differences: As said, in \cite{Cai:2009ac} the author found the $\eta = -{1}/{4}$, in agreement with our (\ref{NR}); see Eq. (2.23) in \cite{Cai:2009ac}. The value $\eta = -{1}/{4}$ was also found in \cite{Hohm:2010jc}; see Eq. (5.70) therein. In order to compare with \cite{Hohm:2010jc} is is necessary to consider that our convention for the sign of the Einstein-Hilbert piece in the gravity action corresponds to $\sigma=+1$ in that paper; besides, they consider conventions with the opposite sign for $m^2$; this is also consistent with (\ref{jjjjj}). In Ref. \cite{Myung:2009up}, the authors find the different value $\eta = 1/16$; see Eq. (27) therein. Another different value appears in \cite{Devecioglu:2010sf}, where $\eta = {7}/{8}$ is obtained; see Eq. (25) therein. In \cite{Gonzalez:2011nz}, the authors found $|\eta |= {1}/{4}$, see Eq. (37) therein, which is actually consistent with our result as they consider the opposite overall sign of the gravity action.

Our result turns out to be consistent with holography. One of the reasons is that it agrees with the result obtained by computing the quasi-local energy with the boundary stress-tensor \cite{Hohm:2010jc}. While in the case of bulk theories whose gravity sector is described by the Einstein-Hilbert action such a computation follows straightforwardly from the holographic renormalization recipe, in the case of higher-derivative theories such as massive 3D gravity the definition of a holographic stress-tensor requires additional prescriptions to define the variational principle and, consequently, to write down the counterterms. This introduces certain degree of ambiguity in the calculation. Therefore, the fact of having reproduced with our computation the results of \cite{Hohm:2010jc} can be regarded as a further support of the definition of the quasi-local stress-tensor proposed therein. Another reason why our result is compatible with holography is that it agrees with the mass spectrum that leads to reproduce the entropy of the Lifshitz black hole from the generalized Cardy formula computation \cite{Gonzalez:2011nz}, which follows from considering the generalization of the modular invariance of the partition function of the dual theory to arbitrary values of $z$. This points into the direction of a microscopic derivation of the Lifshitz black hole entropy.

\[\]

The authors are grateful to L. Donnay, A. Goya and J. Oliva for discussions and collaborations on this subject. The work of G.G. has been supported by CONICET and ANPCyT through grants PIP-1109-2017, PICT-2019-00303.

  \end{document}